\documentclass[twocolumn,showpacs,preprintnumbers,amsmath,amssymb]{revtex4}
\usepackage{graphicx}
\usepackage{dcolumn}
\usepackage{bm}

\newcommand{\dr}{\rightarrow}

\newcommand{\R}{\mathbb{R}}

\def\Ee{{\cal E}}

\def\oo{{\cal O}}

\newcommand{\mone}{^{-1}}
\newcommand{\be}{\begin{equation}}
\newcommand{\ee}{\end{equation}}

\newcommand{\ket}[1]{|#1\rangle}

\def\f{\frac}

\newcommand{\lalg}[1]{\mathfrak{#1}}  

\newcommand{\iso}{\lalg{iso}}

\newcommand{\SO}{\mathrm{SO}}
\newcommand{\ISO}{\mathrm{ISO}}
\newcommand{\arr}{\rightarrow}

\begin{document}
\title{Physics of Deformed Special Relativity:\\ Relativity Principle revisited}
\author{{\bf Florian Girelli\footnote{fgirelli@perimeterinstitute.ca}, Etera R. Livine\footnote{elivine@perimeterinstitute.ca}}}
\affiliation{Perimeter Institute, 31 Caroline Street North Waterloo, Ontario Canada N2L 2Y5}

\begin{abstract}
In many different ways, Deformed Special Relativity (DSR) has been argued to provide an effective limit of  quantum
gravity in almost-flat regime.  Unfortunately DSR is up to now plagued by many conceptual problems (in particular how
it describes macroscopic objects) which forbids a definitive physical interpretation and clear predictions. Here we
propose a consistent framework to interpret DSR. We extend the principle of relativity: the same way that Special
Relativity showed us that the definition of a reference frame requires to specify its speed, we show that DSR implies
that we must also take into account its mass. We further advocate a 5-dimensional point of view on DSR physics and the extension of the kinematical symmetry from the Poincar\'e group to the Poincar\'e-de Sitter group ($\ISO(4,1)$). This leads us to introduce the concept of a pentamomentum and to take into account the renormalization of the DSR
deformation parameter $\kappa$. This allows the resolution of the "soccer ball problem" (definition of
many-particle-states) and provides a physical interpretation of the non-commutativity and non-associativity of the
addition the relativistic quadrimomentum.
\end{abstract}

\maketitle

Quantum Gravity is on the edge of becoming a physical theory. Indeed experiments like GLAST, AUGER and so on, should
measure  effects due to a Quantum Gravity regime. It is important to describe  them in a theoretical framework, and to make predictions. Deformed Special Relativity (DSR) is a good candidate to describe these effects\cite{amelino}. It is mathematically well defined, but its physics is much less understood. Many interpretational problems are plaguing the theory making hard to do clear predictions. In this article we recall what are these problems and we present a new scheme which provides a general solution. First we define the DSR regime as a third regime to be
compared with the Galilean and the relativistic ones. We recall quickly the features and the problems of DSR, before
introducing the new framework. It mainly consists in an extension of the Relativity principle. This comes together with a change of symmetry and thus of the physical objects: particles are now described by a five components momentum and their scattering by the addition of this new pentamomentum. Before concluding,  we present the phenomenology of the new regime. More details on this general scheme can be found in \cite{dsr:gl}.

\section{The DSR regime}\label{regime}

DSR can be considered as a hybrid between Special Relativity (SR) and General Relativity (GR), in which one has imported some gravitational effects, like the notion of the Schwarzschild mass, in the context of SR, i.e. while keeping a flat space-time.

Let us consider a physical object $\oo$, where  $v$, $L$, $M$ are its speed, its characteristic length, its mass (or energy, since there are equivalent in SR) with respect to the reference frame of the observer. General Relativity implies a non-trivial relation between mass and scale coming from the notion of black hole: the Schwarzschild mass provides a maximal mass $M_{max}(L)\equiv \frac{c^2L}{G}$ associated to the scale $L$.
We want to implement this maximum bound in a flat space-time, without breaking the Poincar\'e symmetry, but by
deforming it. This can be interpreted as a (UV) cut-off in momentum space and thus providing a regulator in the context of Quantum Field Theory.

We introduced the DSR regime as describing physics when $M\lesssim M_{max}(L)$. It is defined in contrast with the Galilean regime, $v\ll c$ and $M\ll M_{max}$, and the relativistic regime, $v\sim c$ and $M\ll M_{max}$. 
Note nevertheless that although the DSR regime is defined with no reference to the speed of the objects, DSR effects are enhanced when approaching $c$. Indeed, when $M\arr M_{max}$, we expect gravitational effects to become highly relevant and modify the speed of light. DSR attempts to describe such phenomena from the point of view of an observer ignoring curvature and mapping all physics on his usual flat space-time.

Let us emphasize that the DSR regime is naturally reached when going down to the Planck scale. Indeed, quantum effects induce mass fluctuations $\delta M(L) = \frac{\hbar}{cL}$ depending on the scale of the object (Compton mass). The Planck scale is defined when $\delta M\backsimeq M_{max}$, which is obviously reached when $L$ is the Planck length $L_P$ and therefore $M$ being the Planck mass $M_P$. Thus, at the Planck scale, quantum fluctuations of mass/energy are naturally of the same order of magnitude as the mass bound. This is similar to the well-known argument stating that a high precision position measurement at the Planck scale would directly create a black hole, thus limiting the resolution of measurements of any experiment.

The original motivation for DSR is to postulate a universal maximal mass/energy being the Planck mass, which could be
measured by every observer in any reference frame. The traditional view on DSR is that this universal mass scale becomes a universal mass bound.
Our point of view is that it does indeed make sense to postulate the universality of the Planck
mass -as a signature of the quantum gravity regime- but it doesn't make sense to postulate it is a bound on
energy/mass. Indeed, macroscopic objects have rest energies much larger than the Planck mass. They would hardly make
sense in a theory bounded by $M_P$. This paradox is usually referred in the DSR literature as the {\it soccer ball
problem}. What appears in our simple presentation is that a DSR theory should naturally include a description of the
renormalization of scales and  explain how the bound on energy/mass is renormalized with the scale $L$. Assuming
that general relativity is exactly valid down to the Planck scale, we expect that the mass bound get resized linearly
with the length scale $L$ as expressed above.

Let us consider an observer with resolution the Planck length. He sees the space-time as made of cells of size $L_P$:
the maximal mass $\kappa$ of each object/cell is the Planck mass $M_P=\frac{\hbar}{c L_P}$. Now consider an observer with resolution twice the Planck length. He will see cells of size $2L_P$ and the maximal mass $\kappa$ should be renormalized to $2M_P$. We will prove that DSR induces such a renormalisation. More generally, an object of length scale $L$ in $L_P$ units will be described in DSR with a maximal mass $\kappa\sim \f{\hbar}{cL_P}\frac{L}{L_P}$. Then it is clear that the classical regime of DSR is when $\kappa\dr\infty$, i.e. $L\gg L_P$.

Let us compare DSR with SR: SR introduces a universal speed $c$ which becomes a universal speed bound for all systems, while the DSR deformation parameter $\kappa$ depends on the system. The situation is in fact a bit more subtle. On one hand, SR can be seen as introducing a maximal (Galilean) momentum $mc$ (while the true relativistic momentum remains unbounded) or a minimal energy $mc^2$ (the rest energy), which actually depend on the system (through its mass). On the other hand, DSR can be seen as introducing  a universal length unit $L_P$ and this universal maximal resolution is independent from the observer or the system under consideration and needs to be distinguished from the concept of the mass bound.

In this new DSR regime, we expect new physical features to arise. In Special Relativity,  reference
frames are abstract and don't really correspond to physical objects: reference frames are described through their (relative) speeds with no reference to their mass. In GR, the mass of observers and reference frames becomes relevant due to the gravitational interaction. In the (non relativistic) quantum regime, reference frames can be quantum, i.e. constituted by many particles or only few. It turns out that the mass of a quantum reference frame is highly relevant when dealing with the definition of momenta under changes of reference frame\cite{aharonov}. Or course, these quantum effects disappear for macroscopic reference frames when the mass of the reference frame is very big with respect to the system.
In the DSR regime where we take into account both gravity and quantum effects, it is then natural to assume that the mass of reference frames should be relevant. In this sense one should modify the Relativity principle of Special
Relativity to take into account this mass. Indeed, we will see that changes of reference frames in DSR are not described anymore by a relative speed but by a relative momentum.

Finally, associated to the extra-information of a maximal mass, DSR needs an extended symmetry as well. This is similar to the shift from the Galilean regime and the relativistic regime when we go from the Galilean symmetry group $ISO(3)\times \R$ to the Poincar\'e group $\ISO(3,1)$. To a new symmetry is associated new physical objects, and we will see that DSR is more easily expressed in terms of a pentamomentum which takes into account the mass bound $\kappa$ of the system.

\section{DSR in a seedshell}

In this section we review the basics of DSR and summarize the important problems that have been plaguing the theory. Note that the construction is very similar to the construction of Special Relativity from a deformation of the  Galilean point of view \cite{SR}.  For the most recent updates on DSR, we refer to  the lectures by Kowalski-Glikman, or Amelino-Camelia \cite{amelino}.

The first occurrence of a DSR theory is rather old and now a well-known example of non-commutative geometry. In an attempt to naturally regularize Quantum Field Theory, Snyder introduced in 1947 a theory which incorporates a cut-off in momentum space without breaking  the Lorentz symmetry\cite{snyder}.
He showed that by starting with a non-trivial momenta space,  the de Sitter space, one could retain the Lorentz symmetry, at the price of getting a non-commutative space-time. He considers the momentum as an element of the curved space $SO(4,1)/SO(3,1)$, which can be parameterized using the  five dimensional Minkowski space coordinates $\pi_A$,
\begin{equation}\label{desitter}
-\kappa^2= +\pi_0^2-(\pi^2_1+\pi^2_2+\pi^2_3+\pi^2_4)=\pi^{\mu}\pi_{\mu}-\pi^2_4.
\end{equation}
$\kappa$ is a parameter with dimension of a mass. $\pi_4$ is the fifth direction left invariant under the action of the Lorentz subgroup.  $\SO(4,1)$ acts naturally on the coset space and the Lorentz subgroup generated by the $J_{\mu\nu}$'s acts in the regular way on the five-dimensional Minkowski coordinates $\pi_{A}$, leaving the fifth direction $\pi_4$ invariant:
\begin{equation}\label{desitter1}
\begin{array}{rcl}
[M_i, \pi_j]= i\epsilon_{ijk}\pi_k, &&  [M_i, \pi_0]=[M_i, \pi_4]=0, \\
{[}N_i, \pi_j{]}= \delta_{ij}\pi_0,  && [N_i,\pi_0]= i\pi_i,   {[}N_i, \pi_4{]}=0,
\end{array}
\end{equation}
where we respectively noted as usual $M_i=\epsilon_{ijk}J_{jk}$, $N_{i}= J_{0i}$ the rotations and the boosts.
The four remaining generators of $\SO(4,1)$, which we call the dS boosts, describe the translations on the de Sitter momentum space. 

This deformation of the momentum space is essentially a map from $\R^4$ to the de Sitter space. It
is to be compared with  Special Relativity (SR) as arising as a deformation of the space of speeds,
the space of speeds $\R^3$ being sent onto the hyperboloid $\SO(3,1)/\SO(3)$. There is therefore a strong analogy between the SR case and the Snyder approach to DSR.

Space-time is now reconstructed as the tangent space of the de Sitter momentum space and the coordinates are defined as the (non-commuting) dS boost generators:
\begin{equation}
\begin{array}{rcl}
(X_{i}, X_0)&=& i \frac{\hbar}{c\kappa}( J_{4i},\frac{1}{c}J_{40}),\\
{[}X_{i}, X_{j}{]}&=& i  \frac{\hbar^2}{(c\kappa)^2}J_{ij}.
\end{array}
\end{equation}
The usual relativistic 4-momentum is defined as a choice of coordinate system on de Sitter. Snyder's choice is $p^{\mu}=c\kappa \frac{\pi^{\mu}}{\pi^4}$ and leads to deformed commutators between position and momentum:
\begin{equation}
\begin{array}{rcl}
[X_{i}, p_{j}]&=& i \hbar \left(\delta_{ij} +\frac{1}{(c\kappa)^2}p_{i}p_{j}\right).
\end{array}
\end{equation}

The topic of implementing a maximum quantity with a mass/energy/momentum dimension in a Lorentz invariant setting was then left aside for many years until the phenomenology of both Quantum Gravity and String theory prompted a new interest in the subject, first from the quantum group point of view \cite{ruegg} and then from a more phenomenological point of view \cite{amelino1}.

From the algebraic point of view, the challenge was to introduce a maximal (energy) quantity consistently with the Poincar\'e symmetry. Keeping the generic structure of the Poincar\'e Lie algebra and allowing the deformation of the action of boosts on translations (momenta) while keeping the Lorentz subalgebra untouched, it is possible to show that the set of possible deformations of Poincar\'e is given by the set of solutions of a differential equation. The different solutions provide different sorts of deformations; some bound the energy, others only the 3d momentum, or the rest mass (like Snyder's). Later on, Kowalski-Glikman noticed then that in fact all these algebraic  deformations
can be geometrically understood as different choices of coordinate system on the de Sitter space\cite{kgsnyder} or equivalently as different choices of section for the homogeneous space $\SO(4,1)/\SO(3,1)$. This
important remark made the link between the quantum group approach and Snyder's original approach.
It is important to keep in mind that , by construction, the bounds that are introduced according to the chosen DSR are covariant, i.e. preserved under the Lorentz transformations. There is {\it no breaking of the Lorentz symmetry}.

In this algebraic framework, space-time is then usually reconstructed by duality from the momentum space through the Heisenberg double. More generally, to recover space-time in a DSR theory is nevertheless still an issue as there are different (more or less operational) inequivalent ways to think about the space-time in a non-commutative geometry setting. This is only one among a few deep interpretational problems in DSR, to which we propose solutions:
\begin{itemize}
\item{\bf Multitude of deformations:} The first question which comes up when looking at the
definition of DSR is whether all the different deformations are physically equivalent or not: is there one whichis preferred for physics or does Nature make no difference between them? This is essential as each
deformation singles out a particular new dispersion relation and new conservation laws.

From the algebraic point of view it seems that only one deformation should be physical, and experiments should pick up the only true one. On the other hand, from the geometric point of view, one would be inclined to say that all coordinate systems are equivalent and thus all the deformations should be equivalent.

While these two viewpoints clash, we propose to use the Relativity principle to understand the precise mathematical role of the deformations and check their physical consistency.

\item{\bf Non-commutativity (spectator problem) and non-associativity?}\label{pbnc} To add momenta, i.e. to
build  many body systems, one usually considers the coproduct associated to the algebra of symmetries. For example
in the two particles case, the scattering for the undeformed Poincar\'e is described by the trivial coproduct:
$$
\Delta P= 1\otimes P + P\otimes 1,
$$
which gives the usual addition $p^{(1)}+p^{(2)}$ when applied on a two particles states $\ket{1,2}$. For most of the DSR deformations, the associated coproduct is not (co)commutative.  The non-commutativity naturally induces non-local effects such as the energy of the rest of the universe becoming relevant to the scattering of two particles; this is the  "spectator problem".

Furthermore, most versions of the coproduct are not even (co)associative and thus don't correspond to quantum group like deformations. As an example one can cite the proposal by Magueijo-Smolin\cite{leejoao} who propose a commutative but non-associative coproduct as an attempt to solve the soccer ball problem (see below).

It is true that non-commutativity and non-associativity of the law addition of momenta makes very hard the physical interpretation of physical many-particle states. Our proposal is to interpret this coproduct of the deformed Poincar\'e group as defining composition of momenta for reference frames. The addition of momenta for many-particle states will be later defined with the DSR pentamomentum $\pi_A$. This is to be compared with the situation in SR: the co-product on the hyperboloid defines the law of composition of speeds for reference frames, while the addition of momenta for scattering is the simple commutative and associative addition of the relativistic quadri-momentum.

\item{\bf The soccer ball problem:} The goal to DSR is to introduce a bound in a way which is still compatible the symmetry. The traditional view is to assume that the energy bound $\kappa$ is constant and universal, and to set it to the Planck energy $E_P$. More precisely, assuming that the co-product  of the deformed algebra gives the scattering rule and describes how to build the many-particle states, then we always have the same mass/energy bound for all (many-particle) states. In particular, even macroscopic objects, with rest mass much larger than the Planck mass, should respect the same energy bound $\kappa$ and satisfy the corresponding deformed dispersion relation. This is obviously wrong. Amelino-Camelia coined this problem {\it  the soccer ball problem}. The issue addressed here is more generally how to derive the classical limit of DSR which should describe the classical world with undeformed dispersion relation.

The main idea to solve this problem is to propose that $\kappa$ gets rescaled  with the number of components of
the system. In this sense this is not a constant like the speed of light. Magueijo and Smolin constructed by hand such a scheme\cite{leejoao}: they considered DSR as arising from a non-linear representation of the Lorentz group and introduced by hand in the coproduct a rescaling of the deformation parameter $\kappa$. 
If the intuitive physical origin of this proposition is clear -a renormalization group picture- the resulting
mathematical setting is different than the quantum group approach (or Snyder's) and deserves therefore a
better understanding. Here, we provide solid grounds for the renormalisation of $\kappa$ introducing a natural scattering rule (on the pentamomentum) and showing its compatibility with the Relativity principle.
\end{itemize}

An important point when thinking of DSR physics is that DSR is very likely an effective description of 4d Quantum Gravity (QG) on a flat background \cite{dsrcosmo,dsrgol}. Indeed,  although the theory initially
started as a mathematical trick to regularize Quantum Field Theory, many arguments show it as a possible manifestation of QG. For example, the algebra of observables for one particle in 3d quantum gravity is given by a DSR algebra \cite{dsr3d}.

Here, we attempt to provide DSR with consistent physical foundations and embed it in a clear physical setting, putting the stress on the Relativity principle. Our study, clearly distinguishing the composition of reference frames and the scattering rule and analyzing their compatibility, allows to solve all the conceptual issues mentioned above.

\section{New relativity principle and Pentamomentum} \label{pentamomentum}

In this section, we propose a general framework to consistently interpret DSR. There are two main points. First, reference frames are described by both their relative speed and their mass and we extend the Relativity principle to represent the resulting composition of momenta. Then systems are not solely described by the relativistic quadrimomentum but by a DSR pentamomentum which carries a representation of the dS group and takes into account the mass bound $\kappa$ associated to the system.

\subsection{Reference frames have a speed and a mass}
In Special Relativity, reference frames, physically defined as set of particles, are described only through their relative speeds. Since we are introducing a maximal mass, the mass of a reference frame naturally becomes relevant in DSR. Then reference frames are described by their relative (4-)momentum and not only their relative (3-)speed. As shown in \cite{aharonov}, the mass of a reference frame is already important in the context of usual quantum mechanics. There, a reference frame is made of quantum particles and it is essential to specify its mass for a correct physical description.

Starting with a system whose quadrimomentum is bounded by $\kappa$ in a given reference frame, and is thus represented as a point/vector on the dS space. We would like that the system stays bounded by $\kappa$ under change of reference frame, so that $\kappa$ be a property of the system and do not a priori depend on the particular observer. Similarly to SR where the composition of speeds is represented as translation on the mass-shell hyperboloid, the DSR composition of momentum is represented by translation on the dS space $\SO(4,1)/\SO(3,1)$. {\it Therefore the (non-commutative) co-product of the deformed Poincar\'e algebra describes this composition of momentum under change of reference frames and does not describe the addition of momentum of the scattering rule}.

Moreover, under the choice of a section $p\in dS\hookrightarrow g\in\SO(4,1)$, the composition of momentum as translation on dS comes directly from the group multiplication on $\SO(4,1)$, which now reads as:
\be
g(p_1)g(p_2)=L(p_1,p_2)g(p_1\oplus p_2),
\ee
where $L(p_1,p_2)$ is a Lorentz transformation resulting from the coset structure. A generic choice of section, like Snyder's, will have non-trivial $L$ factors, which lead to the non-associativity of the induced composition of momenta. We call this effect {\it Lorentz precession}.

Our point of view  naturally solves the issue of non-commutativity and non-associativity, and the spectator problem disappears. Indeed, it is not unusual that the composition of reference frames leads to such structures. Already  the composition of speeds of Special Relativity is both non-commutative and non-associative, which become physical features measured through the well-known Thomas precession.

\subsection{Pentamomentum}

DSR proposes an extension of the symmetry group from the Poincar\'e group $\ISO(3,1)$ to the Poincar\'e de Sitter group $\ISO(4,1)$. The natural momentum to consider is now a five-dimensional object $\pi^A$ describing the translations on the 5d Minkowski space. This pentamomentum carries a representation of
the dS group and $\pi^A\pi_A$ is the Casimir of $\ISO(4,1)$. A system is then naturally defined through a generalized mass-shell equation:
\begin{equation}
\pi^A\pi_A=-\kappa^2.
\end{equation}
$\kappa$ still has the dimension of a mass. It is not the relativistic rest mass of the system, and is interpreted as its mass bound. The quadrimomentum $p^{\mu}$ can be computed from $\pi^A$, but its precise expression depends on the particular deformation or equivalently on the coordinate system chosen on dS.

Let us point out an important difference with the SR case:
in SR, the deformation parameter is the speed of light and is universal, while here the deformation parameter $\kappa$ is more similar to the concept of mass in SR and changes with the system. A consequence is that the representation of the symmetry group of the space-time, depending on $\kappa$, changes with the system.

Let us work in the Snyder deformation. The other deformations can be derived from this case by an
adequate coordinate transformation on dS. We define:
\begin{equation}
p^{\mu}=c\kappa\f{\pi^{\mu}}{\pi^4}, \qquad \pi_4^2=\frac{\kappa^2}{1-\frac{1}{(c\kappa)^2}p^{\mu}p_{\mu}}=\kappa^2\Gamma^2.
\end{equation}
This factor $\Gamma$ is similar to the relativistic factor $\gamma^2=(1-v^2/c^2)^{-1}$ of Special Relativity. Indeed, we write $c\pi^{\mu}=\Gamma p^{\mu}$ just as the relativistic momentum reads in term of the Galilean momentum! Note that in the chosen metric $(+----)$, the quadrimomentum of a particle being time-like, we have $1\le\Gamma<\infty$, and the quadrimomentum is bounded in norm by the mass $\kappa^2$: $m^2<\kappa^2$.
Let's point out that the DSR momentum $\pi^A$ is not bounded.

When $ p^{\mu}p_{\mu}\ll \kappa^2$, the DSR momentum coincides with the usual one and we recover the relativistic regime. This is the notion of classical limit. On the other hand when $|p^{\mu}p_{\mu}|\sim \kappa^2$, $\Gamma$ grows arbitrary large and we are fully in the DSR regime.

Just as one interprets $\gamma$ as the new notion of relativistic energy in SR, one is tempted to interpret $\Gamma$ as a new notion of {\it DSR energy}. The main reason to look for a new concept of energy is that the relativistic energy-momentum is not an extensive quantity anymore due to the non-linear deformation of the Poincar\'e algebra: $p_\mu$ is not additive for many bodies, which we would expect for free systems. This is encoded in the non-linearity of the commutation relations (in the Snyder coordinates)
$$
{[}X_i, p_0{]}= i\hbar\f{p_0p_j}{(c\kappa)^2}.
$$
If one considers the DSR pentamomentum instead, it  behaves in the right way with linear commutation relations:
\begin{equation}\label{snyder2}
\begin{array}{lll}
{[}X_i, \pi_4{]}=\frac{i}{\kappa}\pi_i, &
{[}X_i, \pi_j{]}= -\frac{i}{\kappa}\delta_{ij}\pi_4, &
{[}X_i, \pi_0{]}=0, \\
{[}X_0, \pi_4{]}= \frac{i}{\kappa}\pi_0, &
[X_0, \pi_0]= \frac{i}{\kappa}\pi_4, &
[X_0, \pi_i]=0,
\end{array}
\end{equation}
so that it is natural to introduce the DSR energy $\Ee=\pi_4c^2=\Gamma\kappa c^2$. Note that since $\Gamma\ge 1$, even when the rest mass vanishes $m=0$, we still have a non-trivial DSR rest-energy $\Ee_{m=0}=\kappa c^2$.

Let us now describe the addition of pentamomenta. An object is described by
the pentamomentum $\pi^A$ in a given $\kappa$-representation of the Poincar\'e dS group. Let us consider
two-body system made of two objects with the same $\kappa$. The global state of the composite system is described by the diagonal algebra constructed from the product of the two $\iso(4,1)$ algebras of the two objects. We define the global DSR pentamomentum and global Lorentz generators: 
\begin{equation}\label{}
\begin{array}{rcl}
\pi&=&\pi_{(1)}+\pi_{(2)} ,\\
J&=& J_{(1)}+J_{(2)}.
\end{array}
\end{equation}
The coarse-grained position operator $X=\frac{X_{(1)}+X_{(2)}}{2}$ then acts on the new momentum as
\begin{equation}\label{snyder3}
\begin{array}{lll}
[X_i, \pi_j]= \frac{-i\delta_{ij}}{2\kappa}\pi_4, &
[X_i, \pi_0]= 0, &
{[}X_i, \pi_4{]}=\frac{i}{2\kappa}\pi_i, \\
{[}X_0, \pi_0{]}=\frac{i}{2\kappa}\pi_4,  &
[X_0, \pi_i]=0, &
 {[}X_0, \pi_4{]}= \frac{i}{2\kappa}\pi_0.
\end{array}
\end{equation}
Thus, this addition of the pentamomentum directly implies a rescaling of $\kappa$
\begin{equation}
\kappa\rightarrow \kappa'=2\kappa.
\end{equation}
This scaling  for composite systems naturally solves the classical limit problem. The soccer ball problem disappears. For bigger and bigger system, $\kappa$ will grow and we recover the classical Poincar\'e algebra when $\kappa\rightarrow\infty$ since we have:
\begin{equation}
[X^{\mu}, p^{\nu}]= \eta^{\mu\nu}+ O(\frac{p^2}{\kappa^2}).
\end{equation}

Note that because $\kappa$ gets modified, the representation of the operators $p$ and $X$ is modified and the addition is in fact non-linear. Indeed, we have:
$$
p^{j}_{tot}=\kappa '\frac{\pi_1^j+\pi_2^j}{\pi_1^4+\pi_2^4}\neq p_1+p_2=\kappa\,\left( \frac{\pi_1^j}{\pi_1^4}+\frac{\pi_2^j}{\pi_2^4}\right).
$$
This implies that the new representation of the space-time coordinates is
$$\begin{array}{rcl}
X_i&\equiv&i \frac{\hbar}{c\kappa}(\pi^4\partial_{\pi^i}-\pi^i\partial_{\pi^4}) \\&\downarrow& \\
X_i^{tot}&\equiv &i \frac{\hbar}{c\kappa'}\left((\pi^4_1+\pi^4_2)\partial_{(\pi^i_1+\pi^i_2)}-
(\pi^i_1+\pi^i_2)\partial_{(\pi^4_1+\pi^4_2)}\right).
\end{array}
$$
Although $X^{tot}=(X_1+X_2)/2$ at the level of the abstract algebra, this linear relation doesn't hold at the level of their representations in terms of operators only because we are changing the representation and the Hilbert space from $\kappa$ to $2\kappa$.

We can formalize this using the map $U_{\kappa}$ from the 5d Minkowski space to the dS space of curvature $\kappa$:
\be
\begin{array}{rccl}U_{\kappa}:&M^5&\rightarrow & dS_{\kappa}\\
&\pi^A &\rightarrow & p^{\mu}=\kappa\frac{\pi^{\mu}}{\pi^4}.\end{array}
\ee
 Then  $p_{tot}$ reads as:
$$\begin{array}{rcl}
p_{tot}&=& U_{2\kappa}(U_{\kappa}\mone(p_1)+U_{\kappa}\mone(p_2))=U_{2\kappa}(\pi^1+\pi^2)\\
&=&2\kappa\frac{\pi_1+\pi_2}{\pi_1^4+\pi_2^4}=\kappa' \frac{\pi_1+\pi_2}{\pi_1^4+\pi_2^4}.\end{array}
$$
We recognize the  formula proposed by Magueijo and Smolin in \cite{leejoao}. However, unlike them, we are still working in the context of the Snyder deformation and we are not interpreting this addition of momentum as providing a new deformation.  They proposed this non-linearity in order to have a well-defined classical limit. We see that our construction naturally implements this trick, which was also advocated in \cite{dsrgol}.
We provide an explicit construction and more solid grounds for their proposal, showing how this non-trivial addition rule on the $p$'s comes from the trivial addition on the $\pi$'s.

\medskip

More generally, one should deal with composite system made from components of different maximal masses $\kappa_i$.
Our proposal is to add the pentamomenta to get the total pentamomentum. 
The global $\kappa$ of the composite system is the norm of the total pentamomentum. More precisely, one deals with the representation theory of $\ISO(4,1)$ and it is likely we will need to introduce the concept of a "center of maximal mass". The composite system should have then a maximal mass  of the order of magnitude of $\sum \kappa_i$. This construction is exactly the same as when adding objects of different masses in Special Relativity, except that we work with one extra-dimension in the momentum space.
Let us underline some important properties of this scattering rule.
\begin{itemize}

\item{\bf Binding energy:} Let us consider the energy component of total
quadrimomentum resulting from the total pentamomentum. We have
\begin{equation}
E_{tot}=c^2\,\frac{\left(\frac{E_1}{\sqrt{1-(m_1c\kappa)^2}}+
\frac{E_2}{\sqrt{1-(m_2c\kappa)^2}}\right)}{\left(\frac{1}{\sqrt{1-(m_1c\kappa)^2}}+
\frac{1}{\sqrt{1-(m_2c\kappa)^2}}\right)\mone}.
\end{equation}
It is obvious that $\Delta E\equiv E_{tot}-(E_{1}+E_{2})$ does not vanish, and we interpret this difference
as an interaction potential $V\equiv \Delta E$ between the two systems depending on their momenta. This potential forbids the composite momentum to exceed the bound $2\kappa$.

\item{\bf Consistency with the Relativity Principle:} One of the basic postulates of physics is the principle of relativity stating that different observers should still
experiment the same laws of physics. More precisely, we require to have the same laws of conservation in any reference frames. More technically this implies a compatibility relation between the addition of momenta of
the scattering rule and the coproduct describing the composition of momenta under change of reference frames.
Let us consider two systems with pentamomenta $\pi_1, \pi_2$ in a given reference frame. We note $p_1,p_2$ their relativistic momentum. The system $1+2$ has a total pentamomentum $\pi=\pi_1(p_1)+ \pi_2(p_2)$ which is conserved under a scattering process. Let us now make a change of reference frame of momentum $p_{\mu}$. The total pentamomentum in the new reference frame is now $\pi_{p}=\pi_1(p_1\oplus p)+ \pi_2(p_2\oplus p)$. We require that the conservation of the total pentamomentum in all reference frames are equivalent: the conservation of $\pi_p$ must be equivalent to the conservation of $\pi$. Therefore it must be possible  to mathematically express $\pi_p$ in terms of $\pi$ without having to use knowledge about the momentum of the constituents $1,2$: under scattering process, the internal constituents can change and only the total momentum is restrained to be conserved.
One can check that the Snyder deformation actually behaves correctly under this criterium.  More generally, isotropic deformation of Snyder's choice also provide physics consistent with the Relativity principle.

\end{itemize}

\subsection{Deformations: one or many?}\label{soldeformation}
One last big issue is what deformation to use in DSR and whether the deformations are physically equivalent or not.
First, one must be careful not to confuse the actual algebraic deformation which leads to physical consequences and a mere change of coordinates. This is actually the distinction between active and passive coordinate changes (on the dS momentum space).

Now a deformation is a choice of map $f:\R^4\dr dS_{\kappa}$, or equivalently  a choice of section $dS_{\kappa}\hookrightarrow\SO(4,1)$. Isotropic deformations are defined as:
\begin{equation}
\begin{array}{rccl}
f_\varphi:&\R^4&\dr&dS_{\kappa} \\
& p_{\mu}&\dr & e^{i\eta B_{\mu}J^{4\mu}}
\end{array}
\end{equation}
with $p_{\mu}= c\kappa \varphi(\eta)B_\mu$, where $B_\mu$ is of norm one and $\eta$ is the dS boost angle, i.e. the actual distance from the origin on dS. Snyder's deformation corresponds to the choice $\varphi(\eta)=\eta$.

First, one should check whether a deformation leads or not to consistent physics, i.e. consistent with operational principal. Here our criterium is the Relativity principle, which selects the isotropic deformations. Then experimental input is essential to determine the exact deformation $\varphi$. Nevertheless, if we require also linearity of the representation of the $\SO(4,1)$ transformation, then it would select only the Snyder's deformation. However there is no physical principle behind such a choice. This issue is discussed in details in the case of Special Relativity in \cite{SR}. Let us end this discussion by reminding that the coproduct of Snyder's deformation is non-coassociative and thus doesn't correspond to a quantum group deformation.

\section{A New Phenomenology}\label{pheno}

We have proposed a new framework to interpret DSR. The natural question is what is the associated phenomenology. The main point of our proposal is to make the deformation parameter $\kappa$ depend on the system. We have shown that it is consistent with the Relativity principle and that it also a simple solution of the issue of the classical limit of DSR. Nevertheless, it means we have traded all the conceptual issues of DSR for the physical and experimental issue of measuring the parameter $\kappa$ for physical systems. The $\kappa$ is essential in order to make further predictions since it dictates the dispersion relation of the system for example. Our proposal is that $\kappa$ should be the Schwarzschild mass corresponding to the size of the system, i.e. its maximal mass imposed by General Relativity. More generally, the situation is experimentally comparable to Quantum Field Theory where we must first determine the values of the coupling constants before making any further predictions. 

We are  proposing here some physical situations which should provide direct manifestations of the new symmetry. There
are some other physical situations which are usually proposed as sensitive to DSR effects, e.g. GZK cutoff, the
$\gamma$-ray bursts and so on. Those latter should be reconsidered in the context of a field theory expressed in terms of representations of the new symmetry. It is only then that one will be able to make solid physical predictions for these experiments, or more generally the calibration of the deformation.

\begin{itemize}
\item {\bf The Thomas precession}: It indicates the relativistic regime and is easily measurable.
As the action of the Lorentz boosts are deformed and now realized non-linearly, we should have computable corrections to the Thomas precession. This is therefore a new experimental situation that one should explore as a consistency check for DSR.


\item {\bf The Lorentz precession}: Similarly to the Thomas precession of Special Relativity, the law composition of momenta in DSR contains a non-trivial Lorentz transformation depending on the composed momenta. This should have experimental consequences, such as corrections to circular motions depending on the mass of the central object (corrections in nuclear physics?).

\item  {\bf Varying speed of light:} DSR is a well-known example of varying Speed of Light \cite{vsl}. The simple argument is that masses should slow down the motion of light, and  stop it in the extreme case when we reach the maximal mass (black hole case). It is clear that this is still the case in our approach. We expect a redshifting of the very energetic rays of lights i the DSR regime. Such a feature should be seen in  $\gamma$-ray bursts experiments.

\end{itemize}

\section{Outlook}

To sum up the situation, we have proposed a new physical context in which to interpret Deformed Special Relativity. First the deformed co-product on the de Sitter momentum space describes the law of composition of momenta under change of reference frames. In particular, reference frames are not anymore defined only by their relative speeds but one should also take into account their mass and therefore consider their relative momenta. The second step is the shift to an extended Poincar\'e-de Sitter symmetry and the use of a pentamomentum to describe systems. The sum rule for composite systems is to simply add  the pentamomenta of the constituents. A first consequence is that the DSR deformation parameter $\kappa$  is rescaled for bigger systems: it runs to $\infty$ for macroscopic objects and we recover the usual classical world. Then the pentamomentum is conserved under scattering and we have checked that this new conservation law is indeed consistent with the law of change of reference frames: DSR respects the Relativity principle.

For the theoretical point of view, the concept of space-time in DSR is still unclear. The five-dimensional structure of the momentum space and the relation between the bound $\kappa$ and the scale push towards a concept of five-dimensional spacetime where the fifth coordinate is a renormalisation scale. This is similar to ideas expressed in \cite{dsr:spacetime} where 5-dimensional spaces are naturally derived from the renormalisation flow of a 4-dimensional spacetime. The whole point is to understand the physical meaning of the DSR energy and whether there exists an associated notion of time. There is also the standard approach of understanding the operational meaning of spacetime points through the construction of coherent states for fuzzy points \cite{coherent}.

Another issue is whether it is possible to couple the extended relativity principle to an equivalence
principle and derive a deformed general relativity which would take into account the Planck mass.

Finally, the most interesting issue is to study the new DSR phenomenology. It is important to obtain definite physical predictions, on the Thomas or Lorentz precession and the propagation of rays of light, in order to test the theory, and this becomes even more important in the context that DSR provides an effective theory for Quantum Gravity on a flat spacetime.

\section*{Acknowledgments}

We are grateful to Laurent Freidel, Jerzy Kowalski-Glikman, Joao Magueijo, Daniele Oriti and Lee
Smolin, for many discussions on Deformed Special Relativity and for their encouragements in completing our work.


\bibliography{apssamp}

\end{document}